\documentclass{aa}
\usepackage{threeparttable,hhline,longtable}
\usepackage{deluxetable} 
\usepackage{epsfig,latexsym,graphicx,subfigure,verbatim,amsmath}

\usepackage[T1]{fontenc}
\usepackage{ae,aecompl}

\usepackage{natbib}
\newcommand{\cf}{SN~2005cf} 
\newcommand{\coii}{\ion{Co}{\sc{ii}~}}
\newcommand{\feii}{\ion{Fe}{\sc{ii}~}}
\newcommand{\feiii}{\ion{Fe}{\sc{iii}~}}
\newcommand{\siii}{\ion{Si}{\sc{ii}~}}
\newcommand{\siiii}{\ion{Si}{\sc{iii}~}}
\newcommand{\caii}{\ion{Ca}{\sc{ii}~}}
\newcommand{\cii}{\ion{C}{\sc{ii}~}}
\newcommand{\sii}{\ion{S}{\sc{ii}~}}
\newcommand{\mgii}{\ion{Mg}{\sc{ii}~}}
\newcommand{\ia}{{Type~Ia SNe}}
\newcommand{\naa}{\ion{Na}{\sc{i}~}} 
\newcommand{\wl}{$\lambda$}
\newcommand{\sn}[1]{SN~#1}

\newcommand{\kmps}{km s$^{-1}$}                
                
\newcommand{\rsi}{$\mathcal{R}($\ion{Si}{\sc{ii}~}$)$}
\newcommand{\vten}{$v_{10}($\ion{Si}{\sc{ii}~}$)$}    
\newcommand{\dmft}{$\Delta m_{15}$}            
\newcommand{\vdot}{$\dot{v}$}

\begin{document}
 \title{ESC observations of SN 2005cf\\ II. Optical Spectroscopy
   and the high velocity features.}

\author{G. Garavini\inst{1,10},S. Nobili\inst{1,10}, S. Taubenberger\inst{5}, A. Pastorello\inst{5,8},N. Elias--Rosa\inst{2,3,5},  V. Stanishev\inst{1},G.  Blanc\inst{2,9},
S. Benetti\inst{2}, 
A. Goobar\inst{1},
P. A. Mazzali\inst{5,4},
S. F. Sanchez\inst{7},  
M. Salvo\inst{6}, 
B. P. Schmidt\inst{6},
and W. Hillebrandt\inst{5}}

\institute{
Department of Physics, Stockholm University, AlbaNova University Center, SE-10691 Stockholm, Sweden
\and 
INAF Osservatorio Astronomico di Padova, Vicolo dell'Osservatorio 5, I-35122 Padova, Italy
\and 
Universidad de La Laguna, Av. Astrof\'isico Francisco S\'anchez s/n, E-38206 La Laguna, Tenerife, Spain
\and 
INAF Osservatorio Astronomico di Trieste, Via Tiepolo 11, I-34131 Trieste, Italy
\and 
Max-Planck-Institut f\"{u}r Astrophysik,Karl-Schwarzschild-Str. 1, D-85741 Garching bei M\"{u}nchen, Germany
\and
Research School of Astronomy and Astrophysics, Australian National University, Cotter Road, Weston Creek, ACT 2611, Australia
\and 
Calar Alto Observatory Centro Astron\'omico Hispano Alem\'an C/ Jes\'us Durb\'an Rem\'on, 2-2 04004 Almeria, Spain
\and 
Astrophysics Research Centre, School of Mathematics and Physics,
 Queen's University Belfast, BT7 1NN, United Kingdom
\and
APC, Universit\'e Paris 7, 10, rue Alice Domon et L\'onie Duquet, 75205 Paris 
Cedex 13, France
\and LPNHE, CNRS-IN2P3, University of Paris VI \& VII, Paris, France }

 \offprints{G.~Garavini , gabri@physto.se}

 \date{Received ...; accepted ...}
 \authorrunning{Garavini {\em et al.}}

\titlerunning{SN 2005cf}

\date{} \abstract{The ESC-RTN optical spectroscopy data-set for \cf\ is
  presented and analyzed. The observations range from $-11.6$ and $+77.3$
  days with respect to B-band maximum light. The
  evolution of the spectral energy distribution of \cf\ is characterized by the presence of high
  velocity \siii\ and \caii\ features.  SYNOW
  synthetic spectra are used to investigate the ejecta geometry of
  silicon. Based on the synthetic spectra the \siii\ high velocity feature
  appears detached at 19500 \kmps.  We also securely establish the presence of such feature 
in \sn{1990N}, \sn{1994D}, \sn{2002er} and \sn{2003du}.
On a morphological study both
  the \caii\ IR Triplet and H\&K absorption lines of \cf\ show
  high velocity features centered around 24000 \kmps.  When compared with other
  \ia\ based on the scheme presented in \citet{2005ApJ...623.1011B}
  \cf\ definitely belongs to the LVG group.}

\maketitle

\section{Introduction}

Type Ia supernovae (\ia) are the best  standardizable candles currently
known and have proved extremely useful for cosmological studies.  In
the last decade \ia\ have been used to compute the energy budget of
the Universe and to establish the presence of dark energy
\citep{1998Natur.391...51P,1998ApJ...493L..53G,1998ApJ...507...46S,1998AJ....116.1009R,1999ApJ...517..565P,2003ApJ...598..102K,2003ApJ...594....1T,2004ApJ...607..665R,2006A&A...447...31A}.
The observational evidence that brighter/dimmer supernovae have light
curves declining slower/faster \citep{1993ApJ...413L.105P} allows their
use as cosmology lighthouses. Light curve decline rate parameters (for
example \dmft) have been extensively used to divide \ia\ in sub-groups.  Their
observed spectroscopical diversity however, can not be understood
using only one parameter.  Multi-parameter representations of such
diversity based on the spectral morphology \citep{2005ApJ...623.1011B}
or on spectral modeling
\citep{2006PASP..118..560B,2005PASP..117..545B} have been proposed.
These allow an overview of supernova parameter space but fail to be
exhaustive.  Supernova examples such as \sn{1999ac}
\citep{2005AJ....130.2278G} or \sn{2002cx} \citep{2006AJ....132..189J}
do not fit in the proposed classifications and require a deeper
knowledge of supernova physics to be understood.
Careful studies of early SN spectroscopic and polarimetric data have
highlighted that the external structure of the supernova ejecta could be
more complex than once believed. Three-dimensional effects such as
explosion asymmetries and local density inhomogeneity could be at the
origin of the non-spherically symmetric  distribution of the intermediate mass elements
sometimes found inspecting the data.  The discovery of high velocity absorption
features (HVFs) in many of the early supernova spectra recently observed (see
for example \citet{2005ApJ...623L..37M}) have pushed the development
of more complex fully 3-dimensional numerical simulations of 
supernova explosions (see for example \citet{2006A&A...453..203R}). 

In this work we present and analyze the optical spectroscopy data set of
\cf\ acquired by the European Supernova Collaboration  part of the \ia\ Research
Training Network
(ESC-RTN\footnote{http://www.mpa-garching.mpg.de/$\sim$rtn/}). The almost
daily spectroscopic follow-up discloses the evolution of  strong high
velocity features  both in \siii\ and \caii\ absorption lines. This
data set represents a unique opportunity to study the properties of high
velocity features and to investigate their origin.

\section{Observations and data-set} 
SN~2005cf was discovered in the nearby galaxy MCG-01-39-003 
 on a KAIT unfiltered image taken on May 2005, 28.36 UT \citep{2005IAUC.8534....1P}. A
spectrum taken on May 31.22 UT \citep{2005IAUC.8534....3M} showed it
to be a type~Ia SN about ten days before maximum light. Its host
galaxy recession heliocentric velocity is 1937 \kmps (source NED, \citet{1991Sci...254.1667D}).

The ESC extensively followed  SN~2005cf both
photometrically and spectroscopically in the optical and in the near
infrared. The photometry data set is discussed in \citet{pastorello}.   The optical spectroscopy data set presented here  covers the time interval from $-$11.6 days
before B-band maximum light up to 77.3 days after.

All the data (except those observed with the IFU PMAS) were reduced according to the standard procedure.  Using
standard IRAF routines the two-dimensional images were bias-subtracted
and flat-fielded. 
 
Wavelength and flux calibration were applied to the one-dimensional
optimal extracted spectra \citep{1986PASP...98..609H} using calibration observations usually taken with
the same instrumental setting and during the same night as science
observations.  Atmospheric extinction correction was applied via tabulated
extinction coefficients for each telescope used.
PMAS data were reduced using R3D\footnote{http://www.caha.es/sanchez/r3d/}, a package for reducing IFS data \citep{2006AN....327..850S}.

All the spectra were observed at parallactic angle. Synthetic spectrophotometry was computed to adjust the
overall accuracy of the flux calibration to photometric data and no substantial flux correction was required. 
\citet{pastorello} found that the host galaxy reddening was negligible and thus no host galaxy reddening correction was performed.
According to \citet{1998ApJ...500..525S} galactic reddening in the
direction of SN~2005cf is $E(B-V) = 0.097$. i.e. $A_B\sim 0.4$ mag.

\begin{figure*}
\centering
\includegraphics[height=16cm,bb=8 35 565 805,clip]{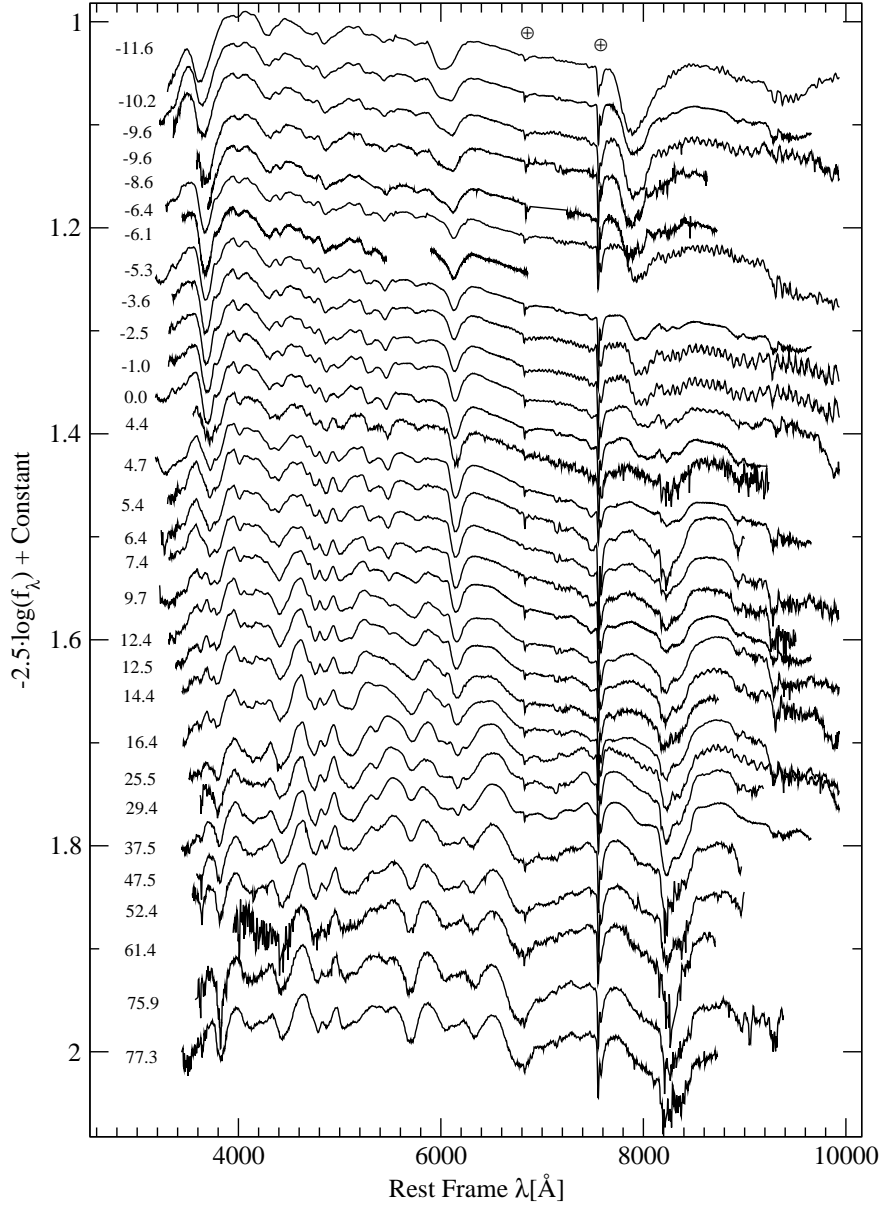}
  \caption{SN~2005cf spectral time sequence. The flux spectra
    are shown in rest frame. The epochs with respect to $B$-band
    maximum light are reported on the left end side of each spectrum. The   $\oplus$  symbol marks the atmospheric absorptions.}
  \label{timeseq}
\end{figure*}

The obtained spectral time sequence is shown in Fig. \ref{timeseq} and
the details of the data set are in Table \ref{tabdataspectra}.
  
\begin{table*}[ht]
\caption{Data set specifications.}
\label{tabdataspectra}
\begin{center}
\begin{tabular}{lrllcllr}
\tableline\tableline
JD & Epoch& Telescope &Instrument&$\lambda$ Range&Exp. Time&Airmass&Resolution\\ 
-2400000   & ref $B_{max}$ & &  &[\AA]  &[sec]&&[\AA]\\
\tableline

53522.4 & -11.6  & TNG     &  DOLORES &3182-9935 &1000&1.3&10\\
53523.8 & -10.2  & ESO-NTT &  EMMI & 3180-9658&1580&1.5&6\\
53524.4 & -9.6  & CA-3.5    &  PMAS & 3587-8950 &1380&1.6&7\\
53524.4 & -9.6  & NOT & ALFOSC & 3357-9933 &900&1.4&13\\
53525.4 & -8.6  & CA-3.5    &  PMAS & 3698-8730 &2000&1.5&6\\
53527.6 & -6.4   & TNG & DOLORES & 3282-9935 &1300&1.3&11\\
53527.9 & -6.1   & SSO-2.3m&  DBSB & 3445-6858 &300&1.5&5\\
53528.7 & -5.3   & ESO-NTT &  EMMI & 3180-9658 &600&1.3&5\\
53530.4 & -3.6   & NOT & ALFOSC & 3344-9934 &900&1.3&13\\
53531.5 & -2.5   & NOT & ALFOSC & 3314-9936 &900&1.3&13\\
53533.0 & -1.0   & NOT & ALFOSC & 3181-9936 &900&1.3&13\\
53534.0 & 0.0   & SSO-2.3m&  DBSB & 3181-9158 &600&1.1&5\\
53538.4 & +4.4   & CA-2.2  & CAFOS  & 3552-9936 &1200&1.4&12\\
53538.7 & +4.7   & ESO-NTT &  EMMI & 3180-9659 &600&1.2&5\\
53539.4 & +5.4   & CA-2.2  & CAFOS  &  3552-9936 &1800&1.4&10\\
53540.4 & +6.4   & CA-2.2  & CAFOS  & 3238-9932 &1800&1.4&10\\
53541.4 & +7.4   & CA-2.2  & CAFOS  & 3314-9932 &1800&1.4&10\\
53543.7 & +9.7   & ESO-NTT &  EMMI & 3215-9658 &600&1.7&4\\
53546.4 & +12.4  & CA-2.2  & CAFOS  &3313-9933  &1800&1.5&12\\
53546.5 & +12.5  & NOT & ALFOSC & 3380-9934 &1200&1.2&13\\
53558.4 & +14.4  & CA-2.2  & CAFOS  &3454-9933 &2400&1.5& 11\\
53550.4 & +16.4  & CA-2.2  & CAFOS  &3454-9933 &2400&1.4&10\\
53559.5 & +25.5  & NOT & ALFOSC & 3510-9936 &1200&1.6&12\\ 
53563.4 & +29.4  & CA-2.2  & CAFOS  & 3430-9935 &1800&1.6&12\\ 
53571.5 & +37.5  & ESO-NTT & EMMI & 3180-9658 &1650&1.1&5\\
53581.4 & +47.4  & CA-2.2  & CAFOS  & 3587-9932 &2700&1.7&10\\
53586.4 & +52.4  & CA-2.2  & CAFOS  & 3545-9932 &2700&2&10\\
53595.4 & +61.4  & CA-2.2  & CAFOS  & 3897-8709 &3600&2.2&12\\
53609.9 & +75.9  & CA-2.2  & CAFOS  &3572-9932 &2700&2.2&12\\
53611.3 & +77.3  & CA-2.2  & CAFOS  &3439-8731 &2700&2.6&13\\
\tableline
\end{tabular}
\normalsize
\end{center}
\end{table*}

\section{Data analysis}
\subsection{Spectral Time Sequence}
The ESC  spectroscopically followed \cf\ for a period of about 3 months starting from 11.6 days before B-band maximum light. Figure \ref{timeseq} shows the complete
time sequence. The first spectrum at day $-$11.6 shows deep absorptions
due to \siii, \mgii,  \feiii, and \caii typical of \ia. From a first
inspection of this spectrum \siii and \caii lines appear to be
exceptionally broad. \siii\ \wl6355 has a flat bottom which develops
to a narrower feature by one week before maximum light. Both the IR
triplet and the H\&K components of \caii\ show broad absorptions
evolving with time toward the appearing of an extra trough on their red
side. The morphological evolution of these absorptions will be
matter of analysis and discussion in the following sections.

A weak \sii\ 'W-shaped' line is visible around 5500 \AA. It strengthens
with time and by day $+$16.4 has completely merged with \sii \wl 5972
and with the strong emerging \naa line forming the characteristic wide
absorption with long and straight blue edge at around 5700 \AA.
The \feii blend absorption feature profile below 5000 $\AA$ at day $-$11.6
is broad and almost featureless. On the red edge of this absorption
two small notches are visible and  develop into a deep trough by
one week prior to maximum light.
From day +25.5 on \cf\ enters the nebular phase and the spectra 
show the typical \feii\ and \coii\ emission lines.

Besides the exceptionally broad \siii\ and \caii\ absorption features
the spectral evolution of \cf\ goes along the line of normal \ia.

\subsection{\ia\ Spectra Comparison}
\label{comparison}

The spectral time sequence of \cf\ (shown in Fig. \ref{timeseq})
highlights the peculiar line profile of some absorption features
which is interesting to compare to those of other normal \ia\ with similar \dmft.  

\begin{figure}
\centering\includegraphics[width=8cm, bb=79 48 691 522,clip]{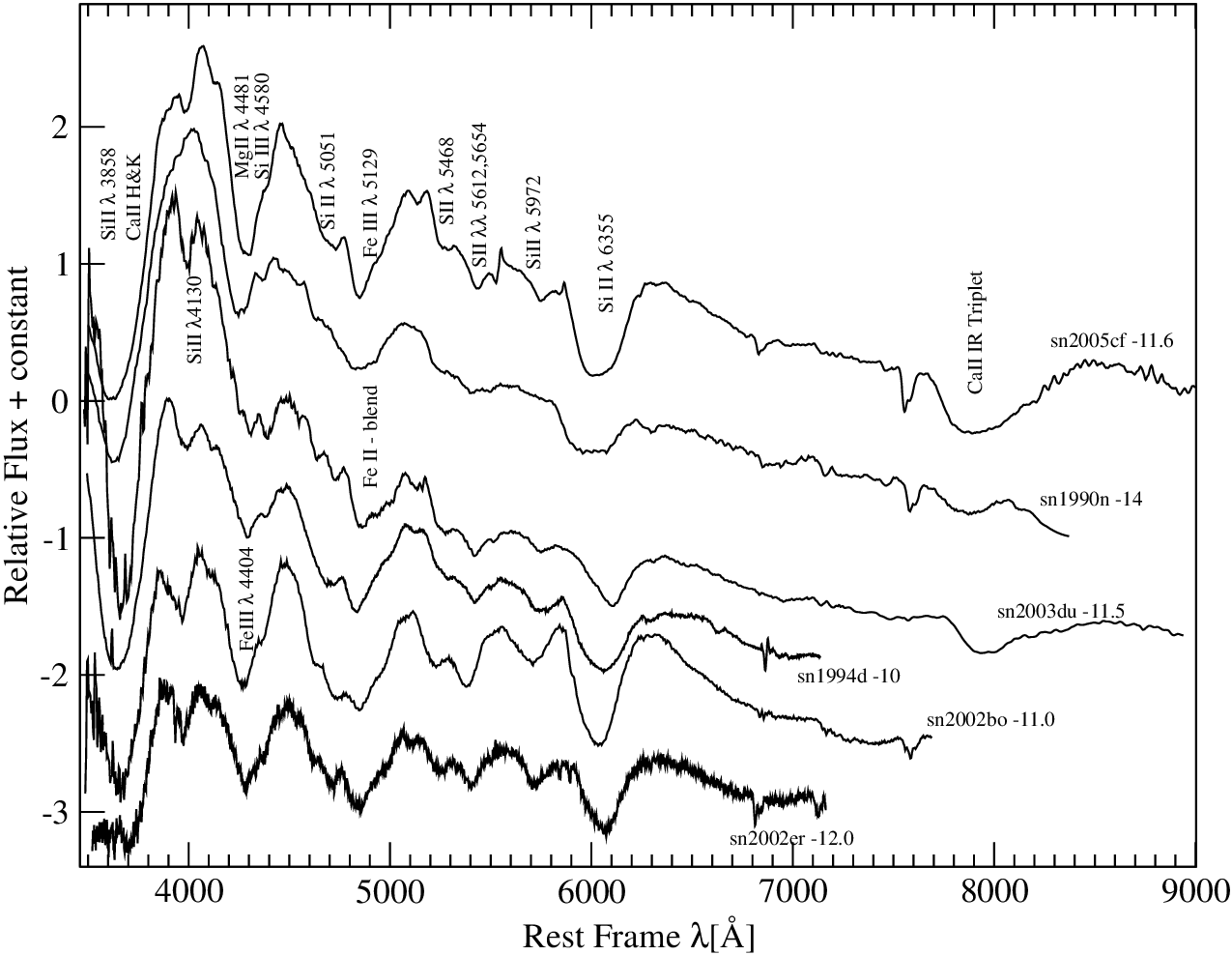}
  \caption{SN~2005cf in comparison with other SNe at day -11.6 (\sn{1990N} \citet{1991ApJ...371L..23L}, 
\sn{1994D} \citet{1996MNRAS.278..111P}, \sn{2003du} \citet{2003du}, \sn{2002bo} \citet{2004MNRAS.348..261B}, \sn{2002er} 
\citet{2005A&A...436.1021K}).}
  \label{comp_-11}
\end{figure}

In Fig. \ref{comp_-11} the  spectrum of \cf\ at day $-$11.6 is compared with
those of other \ia\ at similar epochs. Early time supernova
spectroscopy discloses the physical properties of the ejecta's outermost
layers where \ia\ show the highest degree of heterogeneity. The
spectral absorption features are those of normal \ia\ before maximum
light.  \caii H\&K, and \siii\wl 4130 absorptions have
similar line profile and location as in the other shown SNe. \feiii
\wl 4404, \mgii \wl 4481 and \siiii \wl 4580 form the typical trough
near 4300 \AA. \cf\ and \sn{2002bo} are very similar in this region
while the other shown spectra have different line profiles.
\sn{2003du} has a shallower profile probably due to a weaker
\mgii\ component, \sn{1990N} and \sn{1994D} have a stronger \siiii
line. In the region 4500-5200 \AA\ \cf\ shows a narrower absorption
than the other shown \ia\ with two well distinguished minima due to
\siiii \wl 5051 and \feiii \wl 5129.  This is due to a shallower
\feii-blend component which in the other objects contributes
to making the absorption wider. The \sii 'W'-shaped absorption and the
\siii \wl 5972 are present in all the spectra shown in Fig.
\ref{comp_-11}.  
The main difference in the five spectra compared are around the 6000 
\AA\ and 8000 \AA\ regions.  The \siii \wl 6355 of \cf\ appears broad
 and deep with a flat bottom similar to that of
\sn{1990N}. In \sn{1994D} at 10 days prior to maximum light this
feature is as wide but less deep and without a flat bottom.  The line
profile in \sn{2003du} is different. The minimum of the absorption
feature appears to lie at longer wavelength than in \cf\ but the
trough spans  approximately the same wavelength range. The blue
edge appears less steep. \sn{2002er} and \sn{2002bo} have similar line
profile but different depths.
The \caii triplet region is not covered in all the spectra shown in
Fig. \ref{comp_-11}.  In \cf\ this feature appears to be at shorter
wavelength and it is deeper and wider than in \sn{2003du}.

\begin{figure}
\centering\includegraphics[width=8cm,bb=79 48 679 522,clip]{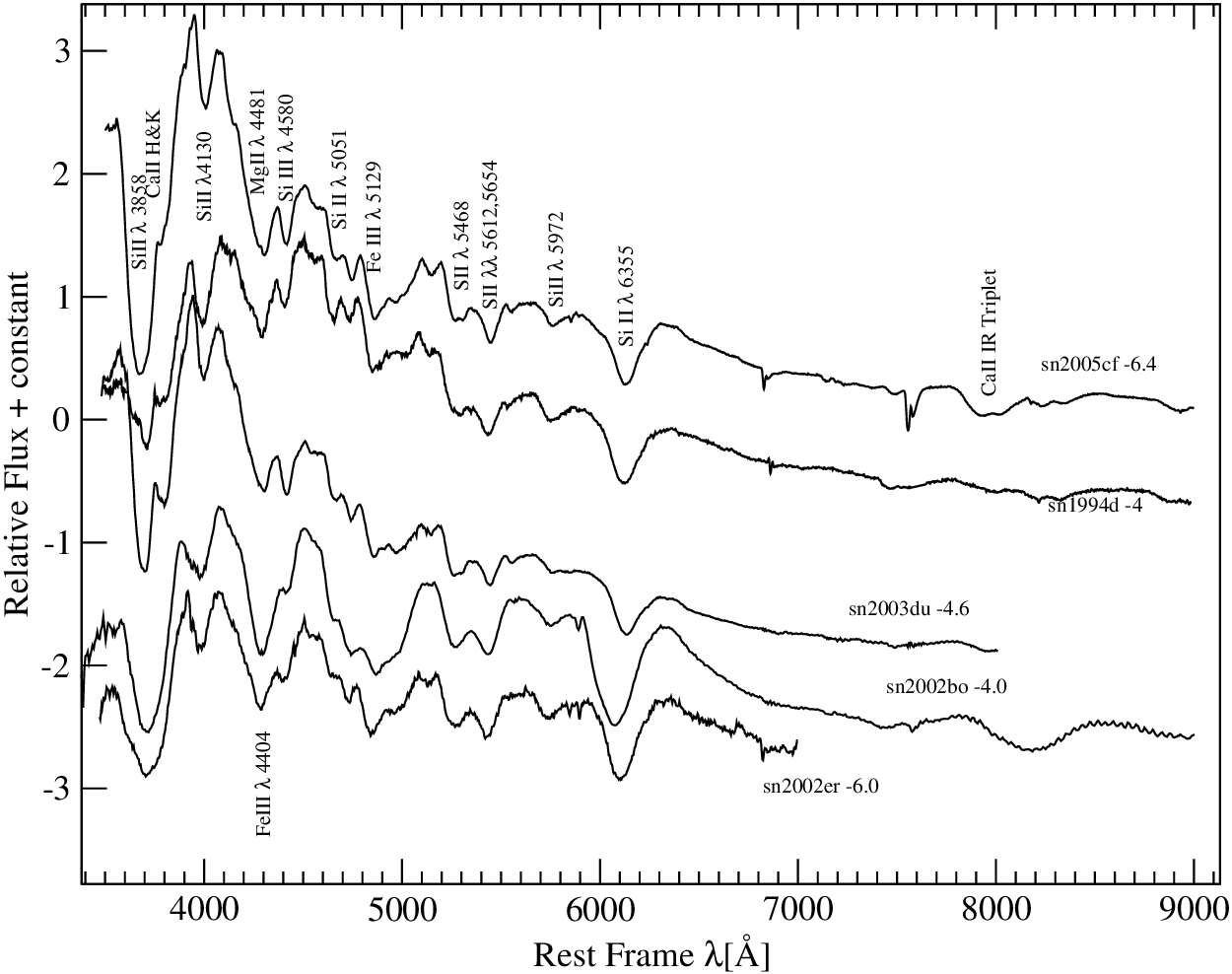}
  \caption{SN~2005cf in comparison with other SNe at day -6.4,
\sn{1994D} \citet{1996MNRAS.278..111P}, \sn{2003du} \citet{2003du}, \sn{2002bo} \citet{2004MNRAS.348..261B}, \sn{2002er}
\citet{2005A&A...436.1021K}).}
  \label{comp_-5}
\end{figure}

 In Fig. \ref{comp_-5} the comparison between the spectrum of \cf\ at about
 $-$6.4 days and those of other SNe is shown. The spectrum of \cf\ has
 evolved maintaining some of the characteristics already shown in the
 earlier epochs. Most of the absorption features developed along the
 lines of normal \ia\ and the spectrum now looks  very similar to that
 of \sn{2003du} at about 5 days before maximum light. The main
 differences stand out on \siii \wl 6355, \caii H\&K and IR
 Triplet. While \caii H\&K on \cf\ has just a hint of a second minimum
 on the red side of the absorption \sn{1994D} and  \sn{2003du} 
 show a well developed one. \sn{2002bo} and \sn{2002er}
 instead, present a broad feature with a single minimum trough. The
 absorption feature around 6100 \AA\ (i.e \siii \wl 6355) in
 \cf\ appears less deep and wide than in \sn{2002bo} and \sn{2002er}, but
 similar to \sn{2003du}. On its blue edge a small notch remains
 visible as residual of the blue component. The \caii\ IR
 Triplet in \cf\ now shows a weaker blue feature which developed a
 double-minimum line profile and a red component which is gaining
 strength. In comparison with the other \ia\ both components appear
 deeper than in \sn{1994D}. \sn{2002bo} instead has  a
 stronger blue but a weaker red component. The spectra of the other
 objects shown in Fig. \ref{comp_-5} do not cover this wavelength
 region.

\begin{figure}
\centering\includegraphics[width=8cm,bb=79 48 686 527,clip]{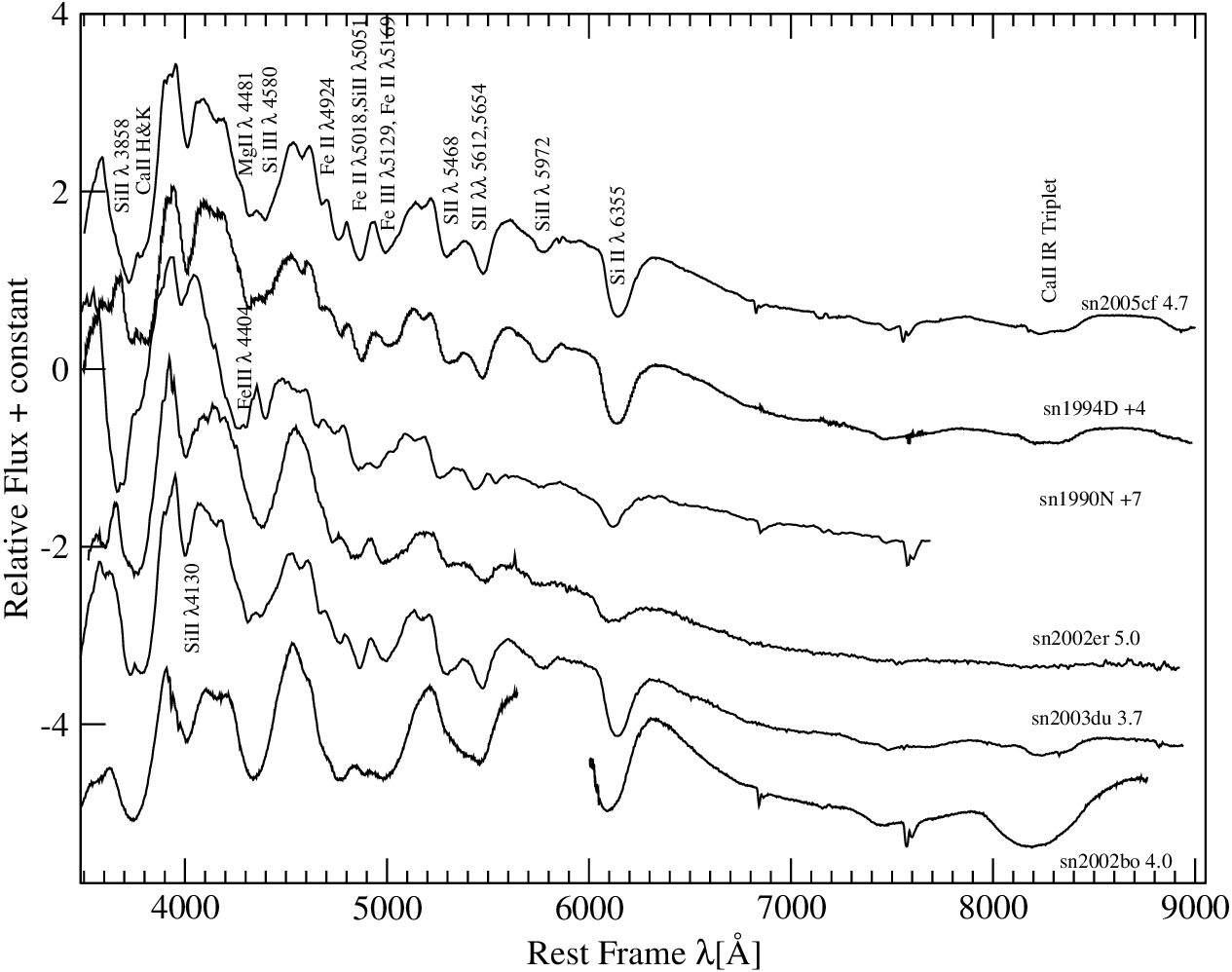}
  \caption{SN~2005cf in comparison with other SNe at day +4.7, (\sn{1990N} \citet{1991ApJ...371L..23L},
\sn{1994D} \citet{1996MNRAS.278..111P}, \sn{2003du} \citet{2003du}, \sn{2002bo} \citet{2004MNRAS.348..261B}, \sn{2002er}
\citet{2005A&A...436.1021K}).}
  \label{comp_+4}
\end{figure}

About 5 days after maximum light the spectra of \cf\ (shown in Fig. \ref{comp_+4}), \sn{2003du}
and \sn{1994D} appear to be very similar while \sn{2002bo} shows wider absorption features all
across the wavelength span  and a deep \caii
IR Triplet at high velocity (i.e on the blue side of the feature).  \sn{2003du} is the closest match to \cf\ in this region. The
absorptions on the blue edge of \siii \wl 6355 and in \caii IR Triplet
have faded away in \cf.

\begin{figure}
\centering\includegraphics[width=8cm,bb=79 48 684 522,clip]{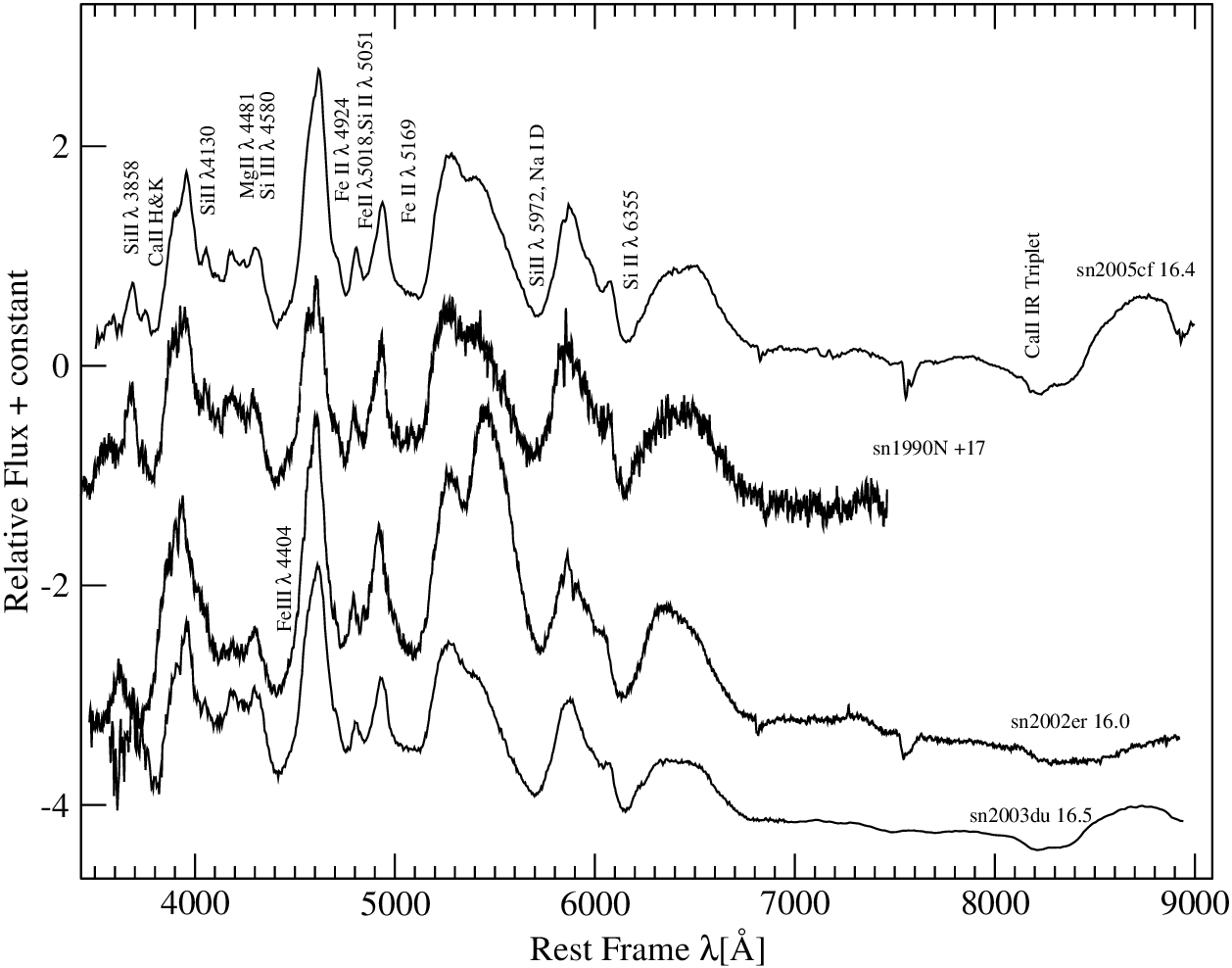}
  \caption{SN~2005cf in comparison with other SNe at day +16.4, (\sn{1990N} \citet{1991ApJ...371L..23L}, \sn{2003du} \citet{2003du}, \sn{2002er}
\citet{2005A&A...436.1021K}).}
  \label{comp_+16}
\end{figure}

At +16.4 days after maximum light \ia\ generally show a smaller degree of
spectral diversity. In Fig. \ref{comp_+16} the compared spectra appear
to be all similar. Stronger \feii\ lines have appeared in the
wavelength region around 5000 \AA\ and the \naa D line now dominates over the
"W"-shaped \sii\ absorption.  
Some degree of diversity is still visible
in the intensity of the \caii\ IR Triplet. \cf\ has a deep absorption
with a pronounced emission component while \sn{2002er} shows a much
weaker feature. \sn{2002er} shows also a unique emission peak  at around 5500 \AA. 

\begin{figure}
\centering\includegraphics[width=8cm,bb=79 48 691 522,clip]{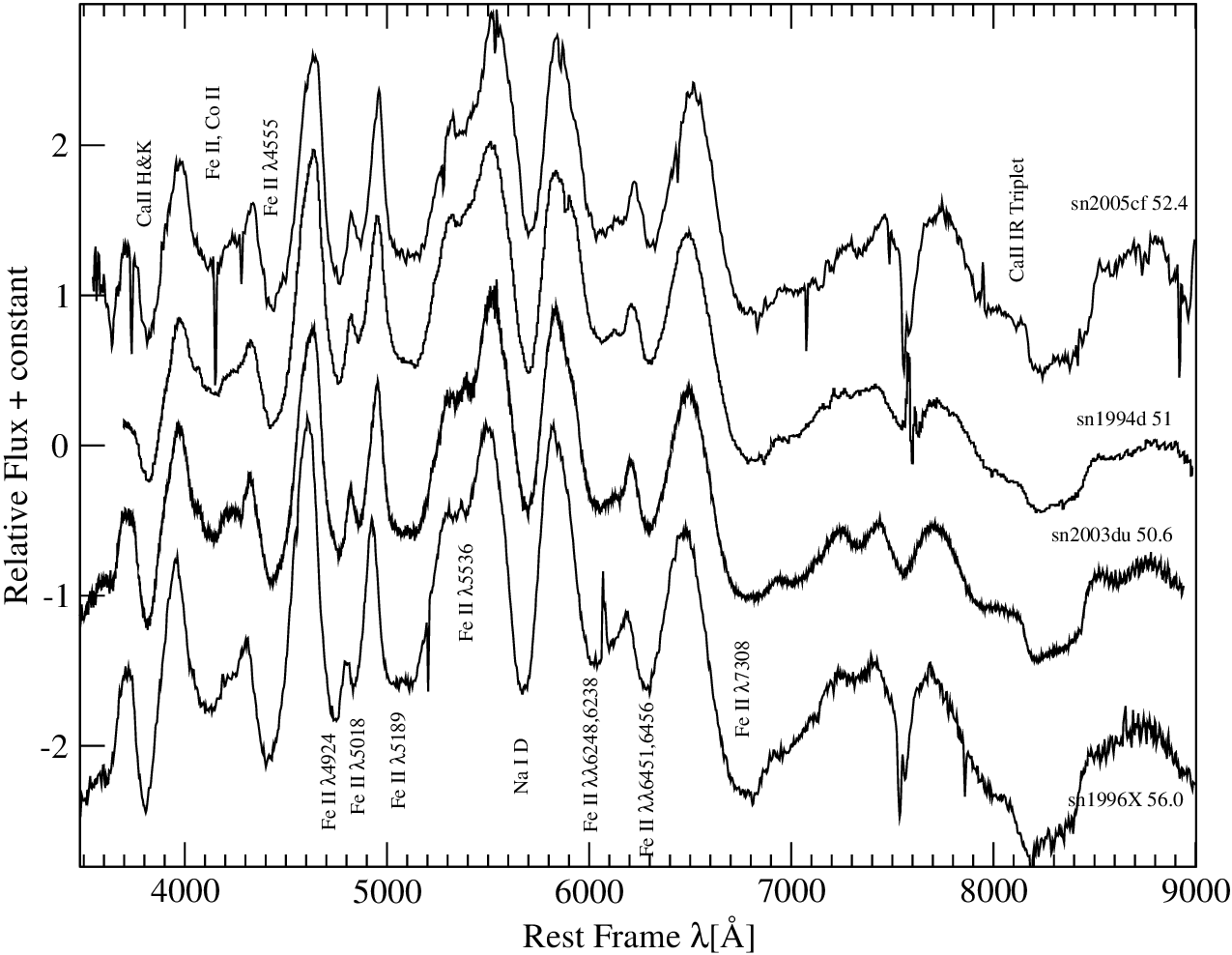}
  \caption{SN~2005cf in comparison with other SNe at day +52.4,
\sn{1994D} \citet{1996MNRAS.278..111P}, \sn{2003du} \citet{2003du}, \sn{1996X} \citet{2001MNRAS.321..254S}.}
  \label{comp_+55}
\end{figure}

In the early nebular phase, at about +52 days, the spectra
compared in Fig. \ref{comp_+55} look very homogeneous. All the spectra
show similar line profiles and intensities in all their emission features.

\subsection{Ejecta Velocity Structure}
The analysis carried out in the previous sections highlights that
\cf\ is overall a normal \ia\ with two peculiarities.  During the
first week of our observations, \siii \wl 6355 appears to be broader
than in most of the other \ia. \caii absorption lines are also broader
and shifted toward bluer wavelength. These characteristics are usually interpreted as
signs of the presence of high velocity material.

\subsubsection{A high velocity silicon feature?}
\begin{figure}
\centering 
\includegraphics[height=8cm,bb= 83 55 281 523,clip]{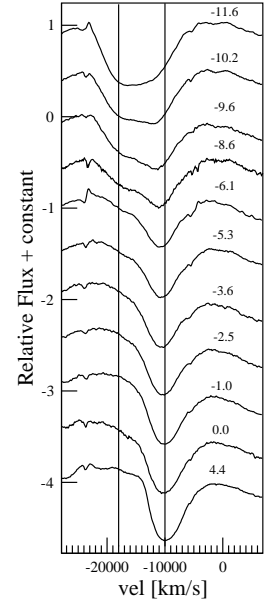}
\caption{The first three weeks of evolution in velocity space of \siii \wl 6355 for
SN~2005cf. Vertical lines indicate 10000 and 18000 \kmps.} 
  \label{siii_vel}
\end{figure}

In Fig. \ref{siii_vel} the time scan of approximately the first
three weeks of spectral evolution of \cf\ is shown, highlighting the
Doppler blueshift of \siii \wl6355. At day $-$11.6 the line profile is
wide and with a flat bottom ranging between 12000 and 18000 \kmps. As the spectra evolve the blue side of the absorption loses
strength and  by day $-3.6$ the feature shows a single minimum at
 about 11000 \kmps. The line profile is still affected
by the blue component  until around maximum light when
\siii\ \wl6355 appears as a single-minimum line at 10000 \kmps  with a steep blue edge.

\label{synow}
An interpretation of the peculiar time-evolution  of the line
profile of \siii\ \wl6355 can be given modeling the spectra with the
use of the parametrized code for supernova synthetic spectroscopy
SYNOW \citep{2000PhDT.........6F}. SYNOW is a direct analysis code
that generates spectra based on a simple, conceptual model of a SN
appropriate during the first few weeks to months after explosion.
This model consists of a blackbody-emitting, sharply defined photosphere surrounded by an extended line-forming, pure scattering
homologously expanding atmosphere.  For each ion introduced, the Sobolev
optical depth as a function of velocity for a ``reference line''
(usually a strong optical line) is specified.  Optical depths in other
lines of the ion are set assuming Boltzmann excitation of the levels
at temperature $T_{exc}$.  The parameters $v_{phot}$ and $T_{bb}$ set
the velocity and blackbody continuum temperature of the photosphere,
respectively.  For each ion, the optical depth $\tau$ at the minimum
ejection velocity $v_{min}$ is specified.  Optical depth scales
exponentially with velocity according to the $e$-folding velocity
parameter $v_{e}$, and is considered to be zero for velocity greater
than $v_{max}$.  If $v_{min} > v_{phot}$ for an ion, we refer to the
ion as ``detached.''

\begin{figure}
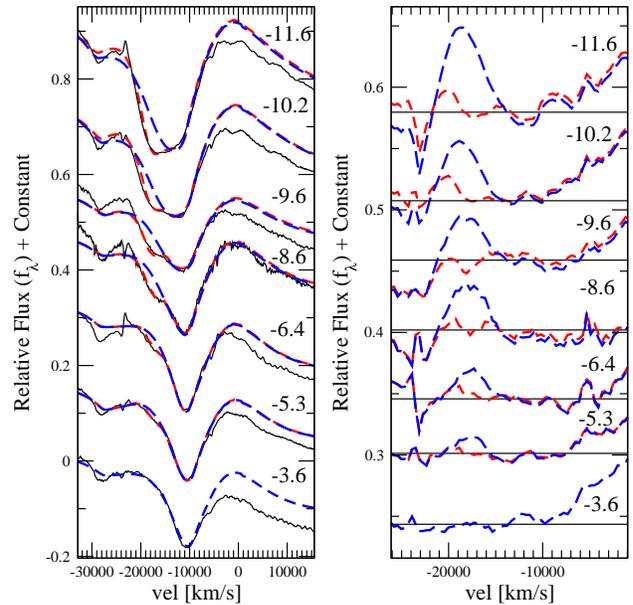

\centering 
\includegraphics[height=8cm,bb=41 55 278 522,clip]{siii_tot_vel.eps}
\includegraphics[height=8cm,bb=41 55 278 522,clip]{res_siii_tot_vel.eps}
  \caption{The evolution in velocity space of \siii \wl 6355 in
SN~2005cf compared with the synthetic models produced with SYNOW ({\it left}) and residuals to the observed spectra ({\it right}).  Long dashed blue lines show SYNOW fits without high
velocity \siii. Short dashed red lines show SYNOW fits with high velocity
\siii.  See Table \ref{synow_table} for details.}
  \label{synow_siii}
  \label{res_synow_siii}
\end{figure}

In our analysis we are interested in constraining the velocity distribution of \siii which appears to be unusual for \ia.
The \siii\ line  visible at about  6150 \AA\ is probably the only SN~Ia feature
unaffected by line blending. To constrain the \siii\ velocity distribution within the ejecta of \cf\  it is thus sufficient to simulate the time evolution of only \siii. 
We then used SYNOW to produce synthetic spectroscopy with only this ion.
In Fig. \ref{synow_siii}  the synthetic spectra are compared with the data obtained during the first week.  Short
dashed red lines show the synthetic spectrum with high velocity \siii,
long dashed blue lines show the synthetic spectrum  without high velocity \siii.  Table \ref{synow_table}
reports the parameters used for the simulation. The HVF appears to be detached at about
19500 \kmps\ with the optical depth decreasing with time. Fig.  \ref{synow_siii} shows
that including a detached high velocity layer to match the HVF in addition to a photospheric velocity layer can accurately reproduce the time evolution
of this region line profile.

\begin{figure}
\centering 
\includegraphics[height=8cm,bb=39 55 292 522,clip]{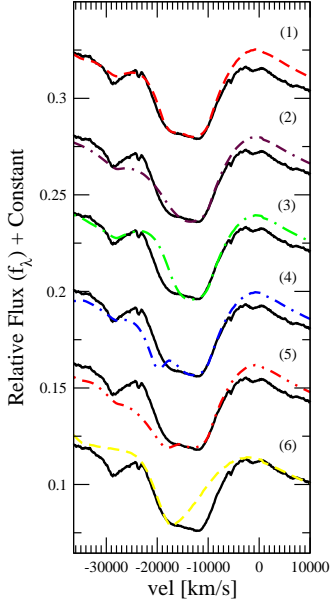}
  \caption{ The velocity space distribution of \siii \wl 6355 in
the $-10.2$ days spectrum of SN~2005cf compared with a variety of synthetic spectra produced with SYNOW. The different synthetic spectra reproduce the effect of varying the key parameters 
through their plausible ranges. (1) Best fit, including  both the HVC and PVC. (2) Fit including only the PVC with $v_{e}=6.3$. (3) Parameters as in the best fit but with the $v_{min}=17.5$ for the HVC. (4) Parameters as in the best fit but with the $v_{min}=21.5$ for the HVC. (5) Parameters as in the best fit but with the $v_={e}=6.3$ for the HVC. (6) Fit including only the HVC detached at  $v_{phot}=21000 $\kmps .}
  \label{synow_more}
\end{figure}

We reached this conclusion by varying the key parameters through their plausible ranges. We reasoned as follows. The observed broad line profile of \siii \wl 6355 could be obtained by simultaneously varying the parameters $v_{e}$, $v_{min}$ and  $v_{phot}$ for a single or a double component of \siii. Figure \ref{synow_more} shows six of the synthetic spectra we obtained varying these parameters (and setting all the others as for the best fit, i.e. fit number 1 in Fig. \ref{synow_more}) while looking for best match the spectrum of \cf\ at $-10.2$ days.
 
The only way to cover the \siii  whole wavelength span with only one component (i.e. the PVC) is to increase its e-folding velocity, $v_{e}$. Fit number $2$ in Fig. \ref{synow_more} 
shows the result of  setting $v_{e}=6.3$ for the PVC. The blue steep profile and the flat bottom can not be accurately reproduce. 
Introducing a HVC helps in reproducing the steep blue edge and to cover the whole wavelength span.  Fits number $3$ and $4$ show the effect of setting the parameter $v_{min}$ for the HVC, respectively to $v_{min}$=17.5 and to $v_{min}21.5$.  Varying $v_{e}$ on the HVC changes the position of the HVC minimum and the profile of its blue edge (as shown in fit number $5$, $v_{e}$=6.3). The best fit (fit number 1 in Fig. \ref{synow_more}) is the result of fine tuning both $v_{min}$ and $v_{e}$ for both \siii components (in addition to matching the optical depths $\tau$).  If PVC is removed and $v_{phot}$ increased  to 21000 \kmps only the blue part of the line profile can be accurately reproduced (fit number $6$).  

\begin{table*}[ht]
\caption{SYNOW parameters for the \siii fits shown in Figure \ref{synow_siii}. PVF indicates the parametres for the photospheric velocity component. HVF indicates the parameters for the high velocity component.}.
\begin{center}
\begin{tabular}{l|lllll|llll|l}

\tableline\tableline
&&&PVF&&&&&HVF&\\
\tableline
day   & $v_{phot}$&$v_{max}$& $\tau$&$T_{exc}$ &$v_{e}$& $v_{min}$& $\tau$&$T_{exc}$ &$v_{e}$&$T_{bb}$\\
  &$10^3$~km$s^{-1}$&$10^3$~km$s^{-1}$&&\small$10^{3}$K &$10^3$~km$s^{-1}$&$10^3$~km$s^{-1}$&&$10^{3}$K&$10^3$~km$s^{-1}$&$10^{3}$K\\
\tableline

-11.6&12&19.5&5.5&10&3.8&19.5&2.5&10&2&11\\
-10.2&11&19.5&6.2&10&3.8&19.5&1.64&10&2.2&11\\
-9.6&10.8&19.5&5&10&3.2&19.5&1.02&10&2&14\\
-8.6&10.8&19.3&3&10&3&19.3&0.5&10&2&9.5\\
-6.4&11.3&19.5&3&10&2&19.5&0.22&10&1&9.7\\
-5.3&10.5&19.5&5&10&2&19.5&0.2&10&1&11\\
-3.6&10.5&-&3&10&2.15&-&-&-&-&11\\

\hline
\end{tabular}
\label{synow_table}
\end{center}
\end{table*}

\begin{figure}
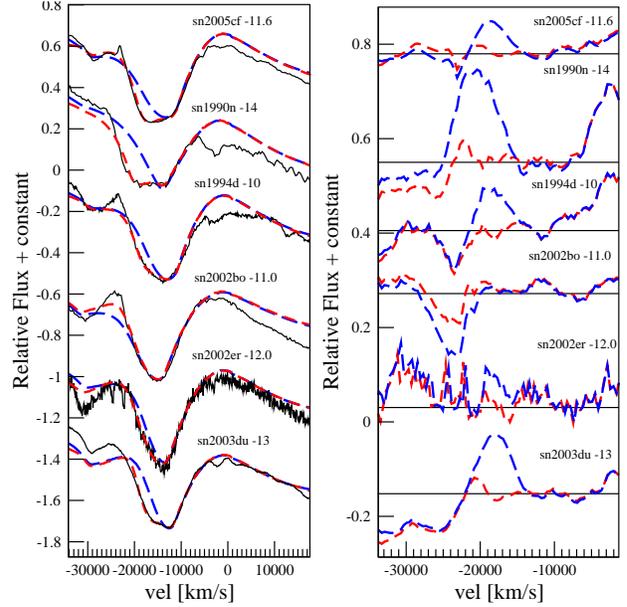

\centering

\includegraphics[width=4cm,bb=75 55 308 522,clip]{xmplot_spec_listm113500_9000_SiIIvel_synow.eps}
\includegraphics[width=4cm,bb=74 55 308 526,clip]{res_xmplot_spec_listm113500_9000_SiIIvel_synow.eps}
\caption{Comparison of the \siii \wl 6355 line profile of different \ia\ at day around $-$11.6 in
velocity space ({\it left}) and residuals to the observed spectra ({\it right}). Long dashed blue lines show SYNOW fits without high
velocity \siii. Short dashed red lines show SYNOW fits with high velocity
\siii. See Table \ref{synow_si.tab} for details.}
  \label{comp_vel_-11}
  \label{res_comp_vel_-11}
\end{figure} 

\begin{figure}
\centering 
\includegraphics[width=4cm,bb=75 55 308 522,clip]{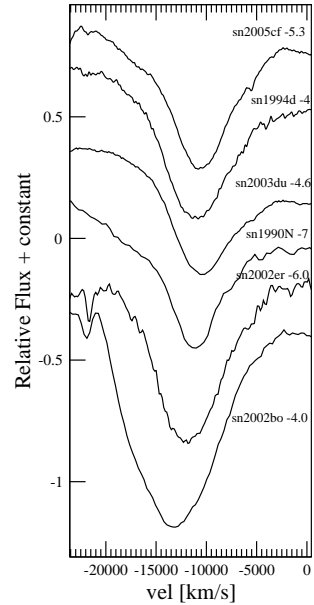}
\caption{ Comparison of the \siii \wl 6355 line profile of different \ia\ at day around -5.3.}
  \label{comp_vel__s-5}
\end{figure} 

The evolution of the relative strength of the two components of
\siii\ in \cf\ appears qualitatively similar to the example reported in
Fig. 12 of \citet{2006astro.ph..3184T}. These authors explored  the effect of different
photosphere covering factors due to high velocity material applyed to a spectrum at maximum light. The same
evolution could be in principle due to the effect of the decreasing
optical depth of a high velocity silicon blob due to the expansion of
the supernova ejecta. In order to validate this hypothesis a detailed
reconstruction of the three dimensional ejecta geometry of the \cf\
 will be carried out in a forthcoming work (Mazzali et al. in
preparation).

\subsubsection{High velocity \siii features in other \ia}

The presence of an HVF in the \siii\ \wl6355 absorption line  has been suggested (see for example \citet{2001MNRAS.321..341M,2005A&A...443..649M,2005MNRAS.357..200M,2003du}) but it could not yet be securely established for other nearby \ia. As we have
seen in the case of \cf\ a daily follow up of the  pre-maximum spectra can highlight the HVF evolution making the identification possible. Inspecting publicly
available early spectra of \ia\  we found a  variety of different \siii \wl
6355 line profiles. We selected the ones that deviated most from the average profile and used SYNOW to simulate their shape.

Figure \ref{comp_vel_-11}  shows the comparison in velocity space of six \ia\ at
day $-11.6$  in the wavelength region
around 6100 \AA\ together with the corresponding SYNOW fits. Short
dashed red lines show the synthetic spectra with high velocity \siii,
long dashed blue lines represent the  syntethic spectra without high velocity \siii. The details of the fits
are reported in Table \ref{synow_si.tab}.

The intensity and position of the high velocity component of
\siii varies among the different \ia.

\begin{table*}
\caption{SYNOW parameters for the \siii fits shown in Figure \ref{comp_vel_-11}. PVF indicates the parametres for the photospheric velocity component. HVF indicates the parameters for the high velocity component.}
\begin{center}
\begin{tabular}{ll|lllll|llll|l}

\tableline\tableline
&&&&PVF&&&&&HVF&\\
\tableline
SN&Day   & $v_{phot}$&$v_{max}$& $\tau$&$T_{exc}$ &$v_{e}$& $v_{min}$& $\tau$&$T_{exc}$ &$v_{e}$&$T_{bb}$\\
  &&$10^3$~km $s^{-1}$&$10^3$~km $s^{-1}$&&\small$10^{3}$K &$10^3$~km $s^{-1}$&$10^3$~km $s^{-1}$&&$10^{3}$K&$10^3$~km $s^{-1}$&$10^{3}$K\\
\tableline

1990N&-14&14.6&21.9&1.15&4&10&21.9&0.72&10&3&11.5\\
1994D&-11&12&20&7&8&3&20&1.2&8&2&12.5\\
2002bo&-11&15&24.5&7.5&10&3&24.5&0.3&10&1&7.5\\
2002er&-12&14&20&7&10&2.4&20&1.1&10&1.8&8\\
2003du&-13&12.8&20&10&10&2&20&2.1&10&1&16\\

\hline
\end{tabular}
\label{synow_si.tab}
\end{center}
\end{table*}

The improvement in the fit achieved introducing a \siii layer at high
velocity is evident in all the cases presented in Fig.
\ref{comp_vel_-11}  with the exception of
\sn{2002bo} for which the contribution of the high velocity component is marginal.
Therefore, we regard the presence of high velocity silicon as highly probable in these spectra.
However, for a positive identification it would be ideal
to densely follow the time evolution of this feature, which requires
daily spectroscopic observations within the first ten days from the
supernova explosion. As it is visible in Fig.~\ref{comp_vel__s-5}, by one week before maximum light the optical
depth of \siii\ in the high velocity layer has dropped to a level
which does not have an evident impact on the line profile of \siii\
\wl 6355. The degree of homogeneity in this line has greatly 
increased by day around $-5$. However, \sn{2002bo} still has a different line
profile due to its overall (i.e in all its spectral features)
exceptionally broad absorpsion features and \sn{1990N} still preserves a hint of the HVF.

\subsubsection{High velocity \caii\ features}

As highlighted by \citet{2005ApJ...623L..37M} the presence of high
velocity features in the  \caii IR triplet is common in many
\ia\ during the pre maximum light phase. In all the cases explored in
\citet{2005ApJ...623L..37M} \caii\ IR triplet HVF loses strength
with time and by around maximum light the absorption at 8000 $\AA$ is
dominated by the photospheric component.

We could expect to see a similar behavior also in the H\&K absorption feature.
 However, this wavelength region is
affected by strong line blending (mainly heavy elements and \siii) and it is
therefore more complicated to surely assess the presence of HVFs by synthetic spectroscopy.

\begin{figure}
\centering\includegraphics[width=8cm,bb=79 52 703 522,clip]{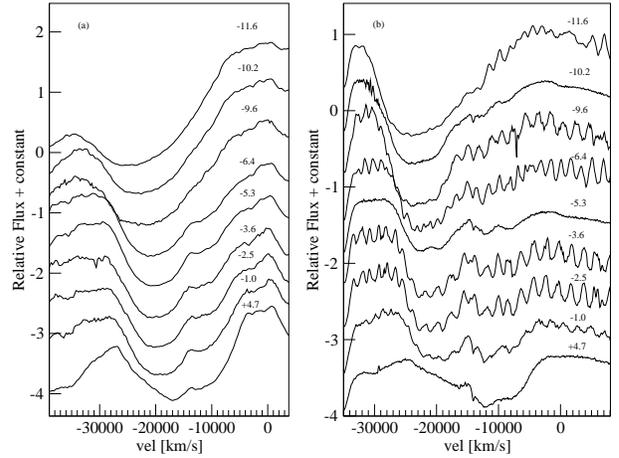}
  \caption{(a) SN~2005cf time scan in velocity space for \caii H\&K. (b) SN~2005cf time scan in velocity space for \caii IR Triplet.}
  \label{hk_vel}
\end{figure}

In the case of \cf\  a simple morphological comparison shows
 similarities between the two absorption features. Figure
\ref{hk_vel} (a) and (b) show the time scan of \caii H\&K
and IR Triplet respectively, between day $-$11.6 and day $+$4.7 in Doppler blue-shift
velocity space. 

In the first spectrum both features show high velocity minima at
about 24000 \kmps. The subsequent evolution of the two
wavelength regions is similar. A second minimum gradually develops at
lower velocity (at about 10000 \kmps) and the corresponding
absorption features strengthens with time.

By maximum light the IR triplet photospheric velocity feature reaches
the same depths as the high velocity one. The same is true one week 
later for \caii H\&K.

\subsubsection{\siii and \caii\ Velocity Time Evolution}

The time evolution of \siii \wl6355 and \caii H\&K  is
interesting to look at in more detail, investigating the expansion
velocity as a function of time at which these absorption features are formed.

\begin{figure}
\centering\includegraphics[width=8cm,bb=21 49 691 508,clip]{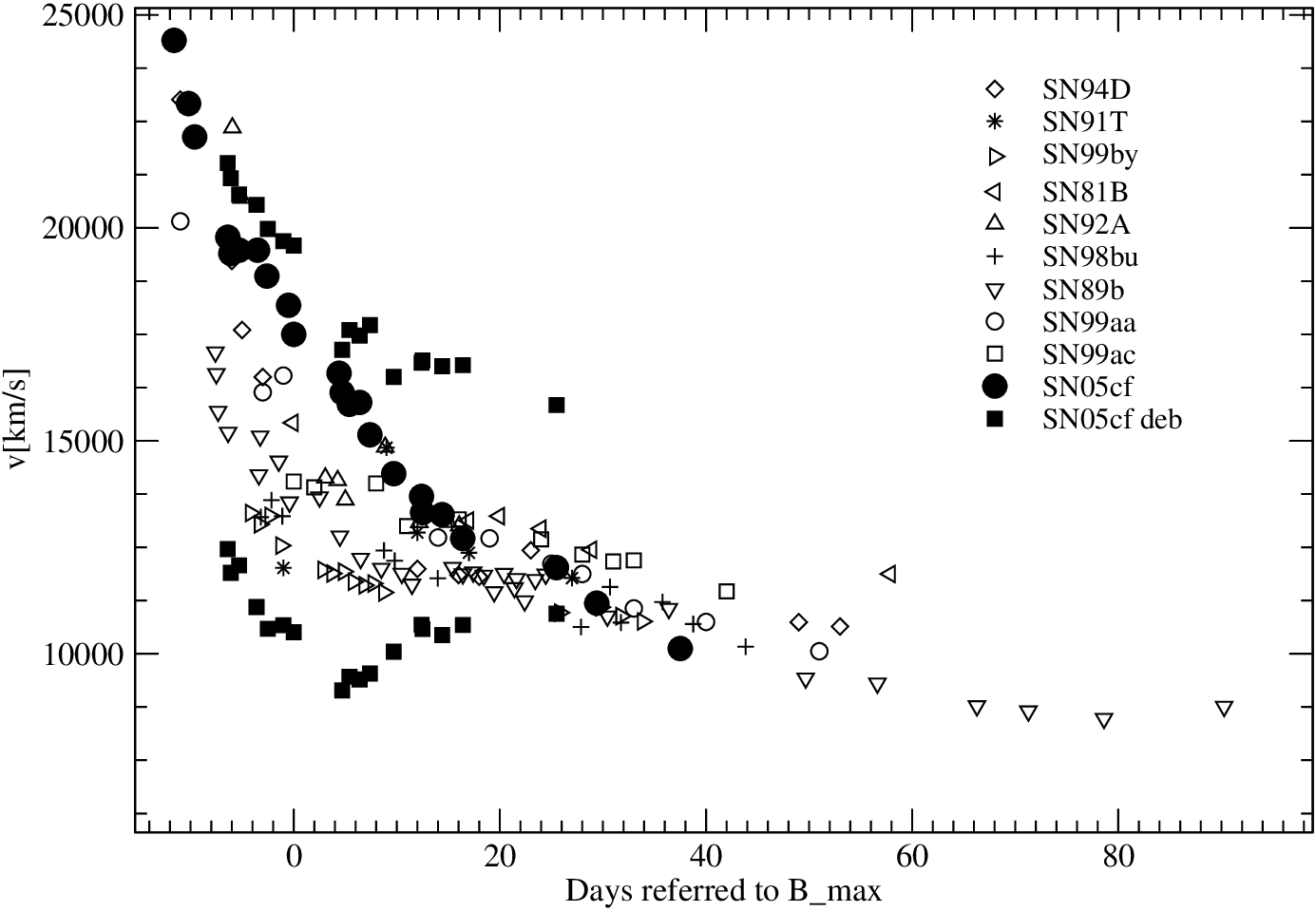}
  \caption{Expansion velocities  as measured from the
  minima of Ca~{\sc ii} H\&K of \cf\ compared with the values of other SNe
  taken from
  \citet{1994AJ....108.2233W,2004ApJ...613.1120G,1993ApJ...415..589K,1996MNRAS.278..111P,1999ApJS..125...73J,2005AJ....130.2278G}
  and references therein. Values for \cf\ are marked as filled
  circles when fitting the whole feature with a single Gaussian model.  Filled squared symbols mark the results obtained when a deblending with a double Gaussian model was performed. Measured values for the single Gaussian model are reported in Table \ref{tabledata}}.
  \label{CaHK_vel}
\end{figure}

Figure \ref{CaHK_vel} shows the \caii H\&K expansion velocity as a
function of time with respect to the B-band maximum light for \cf\ and other \ia.  The plotted
values marked with filled circles are the result of fitting a Gaussian profile to the whole \caii H\&K
absorption line profile.
Between 6 days before and  about 10 days after maximum the
values measured for \cf\ appear higher than the average.
This is due to the persistent presence of the \caii H\&K HVF in  \cf\ which offsets the measured
velocities toward higher values.  At a given phase the relative
strength of the high and low velocity \caii H\&K features for different \ia\
can differ and so can its time evolution. 
In the case of \cf\ the \caii H\&K HVF dominates the \caii H\&K region for
 an unusually long period of time up to one week after maximum light and remains
visible up to approximately three weeks after maximum.
Therefore comparing the velocity measured by fitting the whole \caii H\&K line profile can be
difficult and only qualitative conclusions should be derived from such
comparison.  

 In Fig.  \ref{CaHK_vel}  the filled squares show the result of fitting the  \caii H\&K  profile 
with a double Gaussian model identifying when possible the two visible minima.  These values are however, only indicative of the position of the two  \caii H\&K components. Only an accurate synthetic spectroscopy study of the \caii H\&K  absorption
feature including all the ions contributing to the region could constrain the velocity distribution of the high and low velocity component of \caii H\&K. 

\begin{figure}
\centering\includegraphics[width=8cm,bb=21 49 686 508,clip]{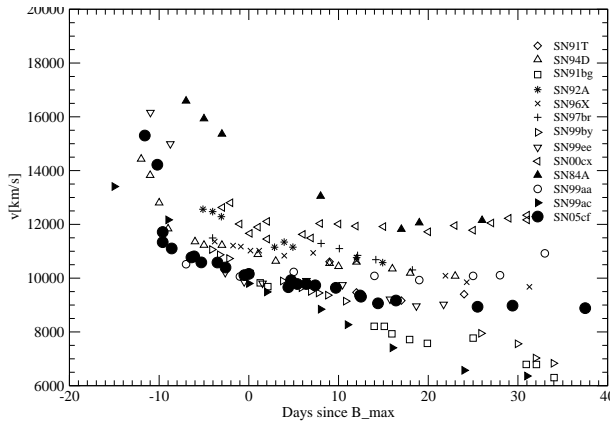}
  \caption{Expansion velocities as measured from the minima of Si~{\sc
      ii}~$\lambda$6355 of \cf\ compared with the values of other
    SNe taken from
    \citet{1999AJ....117.2709L,2001PASP..113.1178L,2004ApJ...613.1120G,2001MNRAS.321..254S,2005AJ....130.2278G}
    and references therein. Values for \cf\ are marked as filled
    circles. Measured values are reported in Table \ref{tabledata}.}
  \label{siII_vel}
\end{figure}

Figure \ref{siII_vel} shows the \siii expansion velocity as a
function of time since the B-band maximum light for \cf\ and other \ia.
The time evolution of the \siii velocity appears to be similar to that
of \sn{1999ee} showing values on the lower edge of normal \ia.

\begin{table}
\caption{Measurements of the expansion velocity (non-relativist Doppler blushift) inferred from
\caii~H\&K and \siii~$\lambda$6355 for \cf.}. 
\label{tabledata}
\begin{center}
\begin{tabular}{ccc}
\tableline\tableline
Epoch&\caii~H\&K&\siii\\
 days&[\kmps]&[\kmps]\\
\hline
           -11.6  & 24403  & 15304   \\
           -10.2  & 22920  & 14218   \\
            -9.6  & --     & 11339   \\
            -9.6  & 22137  & 11716    \\ 
            -8.6  & --     & 11103   \\
            -6.4  & 19780  & 10772   \\  
            -6.1  & 19400  & 10810   \\
            -5.3  & 19476  & 10585   \\
            -3.5  & 19476  & 10583    \\ 
            -2.6  & 18867  & 10395    \\
            -0.5  & 18183  & 10111    \\ 
	     -0.  & 17499  & 10159    \\ 
             4.4  & 16586  & 9663     \\
             4.7  & 16130  & 9923     \\
             5.4  & 15856  & 9781     \\
             6.4  & 15902  & 9781     \\
             7.4  & 15141  & 9734     \\
             9.7  & 14229  & 9639     \\
            12.4  & 13696  & 9356     \\
            12.5  & 13316  & 9309     \\ 
            14.4  & 13267  & 9064      \\
            16.4  & 12708  &  9167   \\
            25.5  & 12023  &  8936    \\
            29.4  & 11187  &  8978    \\
            37.5  & 10122  &  8884    \\
\hline
\end{tabular}
\end{center}
\end{table}

\subsection{High Velocity Features Possible Interpretation}
\citet{2004AJ....128..387G} have shown that the presence of HVFs 
of \caii , \cii and \feii can be reproduced via an
artificial detached line strength in the framework of SYNOW. In the present work we have
produced evidence that  in \cf\ as well as in other \ia\ the same is true for \siii.

The physical interpretation of such high velocity absorption features in terms of
the supernova progenitor system evolution and explosion mechanism and
dynamics is still under investigation.

The additional line strength in the external layer of the ejecta
can be interpreted as abundance or density enhancements.
However, \citet{2005MNRAS.357..200M} have shown  that it is not possible
to reproduce strong HVFs just letting Si and Ca dominate the outermost
layers of the supernova ejecta. Density enhancements must  also be 
involved. Three-dimensional explosion models show  that
it is possible to create local density enhancements in the form of
blobs of incomplete silicon burning material at high velocity \citep{2006A&A...453..203R}. This
scenario is supported by the strong  polarization of some of the HVFs
observed in some \ia, see for example \citet{2003ApJ...591.1110W}.
Alternatively \citet{2006ApJ...636..400Q} and \citet{2004ApJ...607..391G} proposed that high
velocity density enhancement could be the result of an interaction of
the ejecta with a circumstellar shell of solar composition.

To disentangle between these different scenarios or understanding their
interplay is beyond the scope of this analysis.
Eventually,  only  daily
spectropolarimetric follow up observations from the very first days of
the explosion will  provide the necessary data set to disclose the origin of HVFs.

\subsection{Supernova Parameter Space}

Type Ia supernovae are currently believed to be a multi-parameter class
of objects.  Several parameters have been proposed with the goal of
exploring \ia\ diversity. However, the exact number of parameters needed to
describe the observational evidence is however not yet known.

In the following we make use of the parametrization proposed by
\citet{2005ApJ...623.1011B} to quantitatively compare \cf\ with other
observed \ia. 

The parameters used for such comparison are:
\rsi\ \citep{1995ApJ...455L.147N}, \vten\ \citep{1993AJ....105.2231B}, \vdot\ \citep{2005ApJ...623.1011B} and \dmft \citep{1993ApJ...413L.105P}.  The measured values for \cf\ are
reported in Table \ref{values}.  The symbols used in the following
figures are chosen to match  those in \citet{2005ApJ...623.1011B} to
maintain the same distinction among the three clusters found in their
analysis. According to this parameterization \ia\ can be divided in
three groups (see \citet{2005ApJ...623.1011B} for details): FAINT (low
luminosity and high \dmft), HVG (high velocity gradient \vdot) and
LVG (low velocity gradient \vdot).

\begin{figure}
\centering\includegraphics[width=8cm,bb=54 360 558 720,clip]{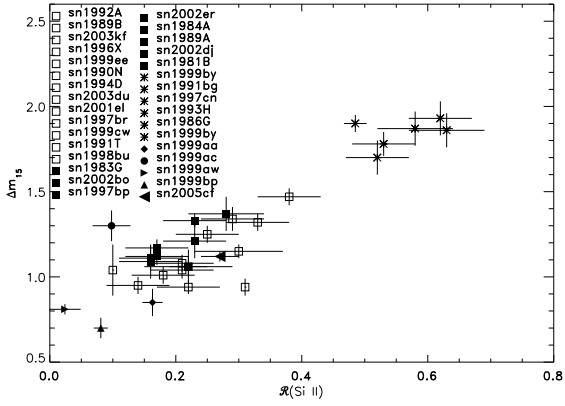}
  \caption{Light curve decline rate \dmft\ versus \rsi.  The symbols are chosen accordingly with
    in \citet{2005ApJ...623.1011B}. Open squares mark SNe belonging to the LVG group, filled squares mark the HVG group and starred symbols mark the FAINT group. Other filled symbols mark the objects presented in \cite{2005AJ....130.2278G} and \cf.  }
  \label{siratio}
\end{figure} 

In Fig. \ref{siratio} the value of \dmft\ versus \rsi\ is shown in
comparison to other \ia. According to these parameters \cf\ could be
identified belonging to both the HVG or LVG groups.

\begin{figure}
\centering\includegraphics[width=8cm,bb=54 360 558 720,clip]{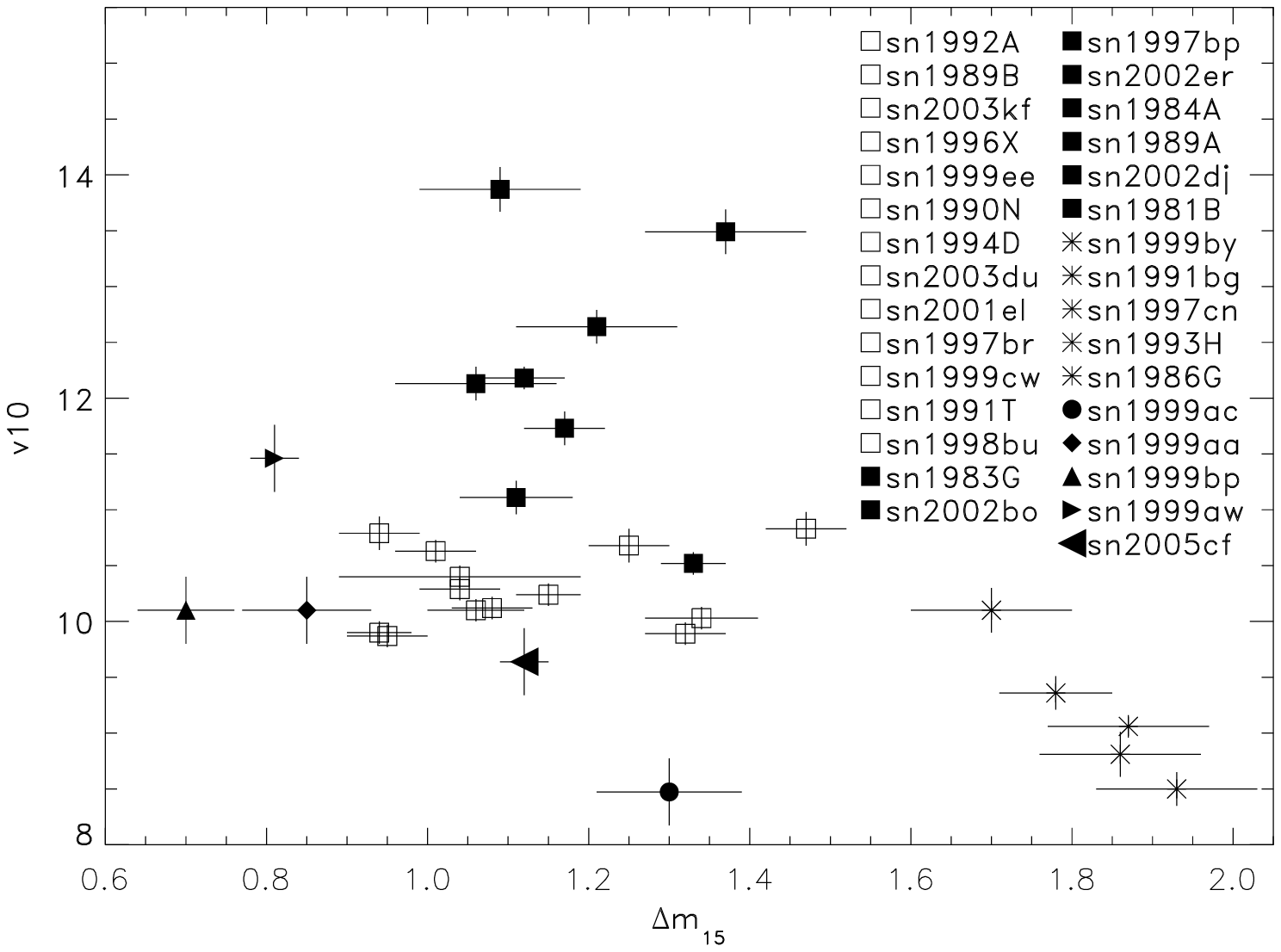}
  \caption{Plot of \dmft\ versus \vten.  Symbols as in Fig. \ref{siratio}}
  \label{dm15vsv10}
\end{figure} 

The comparison  (shown in Fig.  \ref{dm15vsv10}) between the values of
the light curve decline rate and the \siii\ velocity at 10 days past
maximum light (i.e \vten)   positions \cf\ at the lower edge of the LVG
group showing a lower velocity value.

\begin{figure}
\centering\includegraphics[width=8cm,bb=54 360 558 720,clip]{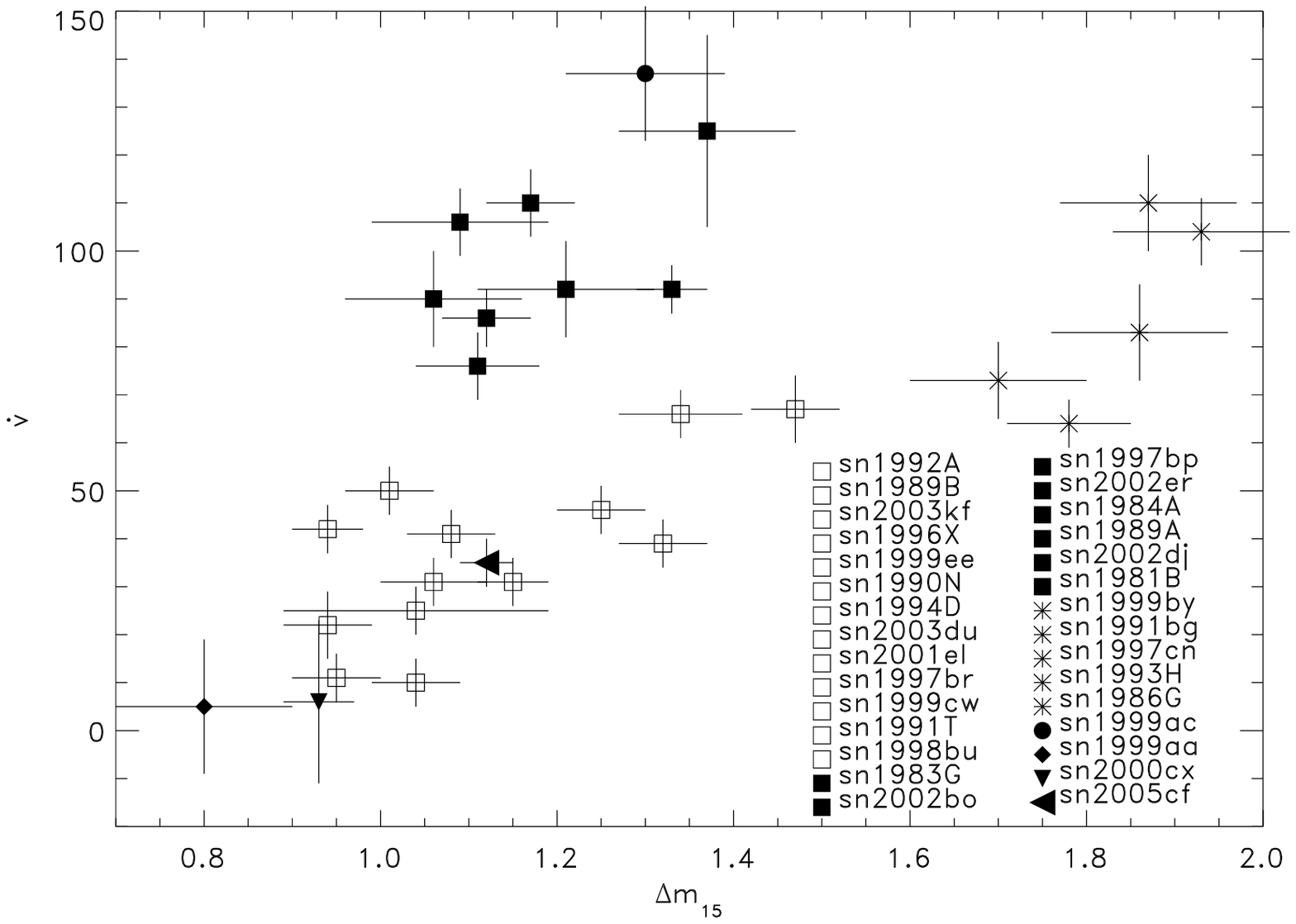}
  \caption{Plot of \dmft\ versus \vdot.  Symbols as in Fig. \ref{siratio} }
  \label{dm15vsvdot}
\end{figure} 
Also in the plane \dmft\ versus \vdot\ (Fig. \ref{dm15vsvdot})
\cf\ belongs to  LVG group showing  a low velocity
gradient.

\begin{figure}
\centering\includegraphics[width=8cm,bb=54 360 558 720,clip]{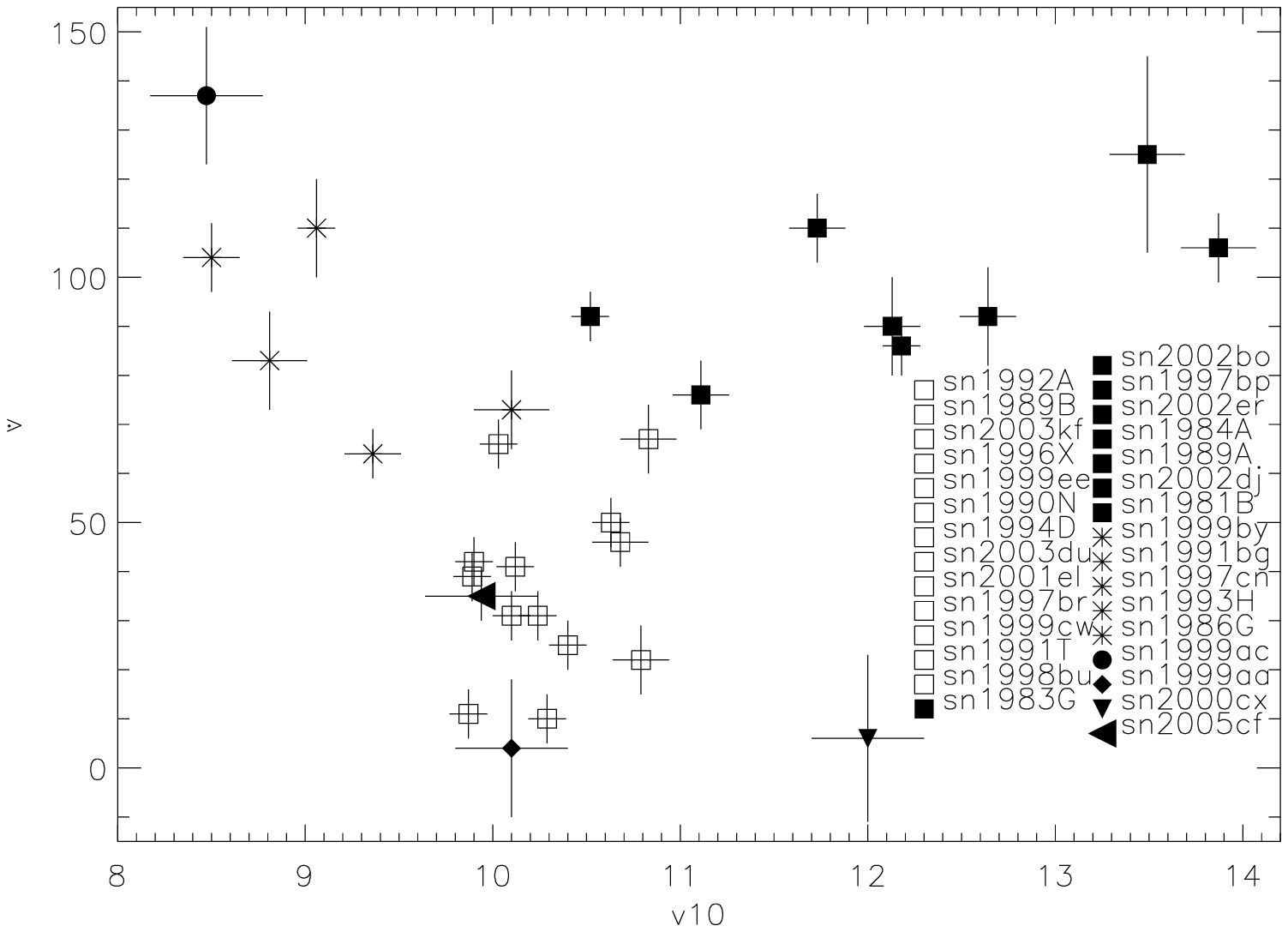}
  \caption{Plot of \vdot\ versus \vten.  Symbols as in Fig \ref{siratio} }
  \label{vdotvsv10}
\end{figure} 

In the \vdot\ versus \vten\ plane (Fig.  \ref{vdotvsv10}) \cf\ seems to
 belong the LVG group
showing a low velocity gradient value as well as a low silicon velocity
at 10 days past maximum light.

In conclusion in the classification scheme identified in \citet{2005ApJ...623.1011B}
\cf\ is definitively a LVG SN~Ia. The presence of the HVFs does not seem to affect the classification scheme defined in  \citet{2005ApJ...623.1011B} since it is mostly defined trough the properties of 
spectra at or after maximum, when the HVFs are less important. 

\begin{table}
\caption{Measured values of RJD$_{B}^{max}$, M$_{B}^{max}$, \dmft, \rsi, \vten, \vdot\
for \cf. Errors are indicated in parenthesis.\label{values}}
\begin{center}
\begin{tabular}{ll}
\tableline
\hline
RJD$_{B}^{max}$&53534.0\tablenotemark{*}\\
M$_{B}^{max}$&-19.39 (0.33)\tablenotemark{*}\\
\dmft&1.12 (0.03)\tablenotemark{*}\\
\rsi &0.27 (0.03)\\
\vten&9939 (300)\\
\vdot&35 (5)\\

\hline
\end{tabular}
\tablenotetext{*}{As measure in \citet{pastorello}}
\end{center}
\end{table}

\section{Summary and Conclusions}
We have presented the optical spectroscopy of \cf. The data collected
by the ESC-RTN collaboration range from $-11.6$ to $+77.3$ days with
respect to B-band maximum light. Based on the analysis presented in
this paper \cf\ appears to be a normal \ia\ with two interesting
peculiarities. 

The early supernova spectra show the presence of high
velocity features of both \caii IR Triplet and H\&K detached around
24000 \kmps. The appearance of a HVF in the \caii IR Triplet is common to
many \ia\ \citep{2005ApJ...623L..37M} but the corresponding visibility
of a \caii H\&K component at high velocity is not equally widespread.

The second rare characteristic is the clear presence of a detached
layer or blob of \siii\ at about 19500 \kmps. We establish the occurrence of
the corresponding high velocity features by means of SYNOW synthetic spectra. 

Moreover, we also securely identify the presence of such feature in \sn{1990N}, \sn{1994D}, \sn{2002er} and \sn{2003du}. 
This shows that the presence of high velocity \siii\  features is not uncommon in early \ia spectra. However, in order to observe its time 
evolution, daily early time spectroscopic follow up must be acquired.

We carry out a quantitative comparison between \cf\ and other \ia\ based on the
parameterization presented by \citep{2005ApJ...623.1011B}. \cf\ appears to belong to the LVG group, showing a slow expansion
velocity at 10 days after maximum light as well as a low velocity gradient. 

Abundance or density enhancements, and  the interaction of the
expanding ejecta with a circumstellar shell with solar composition have
been proposed to interpret the peculiar ejecta structure found
in \ia\ similar to \cf.  Disentangle among these explanations will
shed light on the \ia\ explosion physics and will be the topic of a
forthcoming work.

\begin{acknowledgements}
This work has been partially supported by the European Union's Human
Potential Programme ``The Physics of Type Ia Supernovae'', 
under contract HPRN-CT-2002-00303.
G.G, A.G. and V.S. would also like to thank the
G\"{o}ran Gustafsson Foundation for financial support.\\
This paper is based on observations collected at the
Centro Astron\'omico Hispano Alem\'an (Calar Alto, Spain), Siding Spring Observatory (Australia),  Telescopio Nazionale Galileo, Nordic Optical Telescope (La Palma, Spain), ESO-NTT Telescope (La
Silla, Chile).  We thank the support astronomers for performing the observations.
We thank Thomas Augusteijn, Eija Laurikainen, Karri Muinonen and Pasi
Hakala for giving up part of their time at the Nordic Optical Telescope
(NOT), and Jyri N\"ar\"anen, Thomas Augusteijn, Heiki Salo, Panu Muhli,
Tapio Pursimo, Kalle Torstensson and Danka Parafcz for performing the
observations. We are also grateful to P. Sackett
for the help in observing SN~2005cf from Siding Spring.\\
This research has made use of the NASA/IPAC Extragalactic
Database (NED) which is operated by the Jet Propulsion Laboratory,
California Institute of Technology, under contract with the National
Aeronautics and Space Administration. We also made use of the Lyon-Meudon
Extragalactic Database (LEDA), supplied by the LEDA team at
the Centre de Recherche Astronomique de Lyon, Observatoire de Lyon.\\

\end{acknowledgements}

\bibliography{2005cf.bbl}

\end{document}